\documentclass[runningheads]{llncs}

\usepackage{graphicx}
\usepackage{amsmath}
\usepackage{amssymb}
\usepackage{booktabs}
\usepackage{subcaption}

\usepackage[pagebackref,breaklinks,colorlinks]{hyperref}

\usepackage[capitalize]{cleveref}
\crefname{section}{Sec.}{Secs.}
\Crefname{section}{Section}{Sections}
\Crefname{table}{Table}{Tables}
\crefname{table}{Tab.}{Tabs.}

\begin{document}

\title{A Hierarchical Descriptor Framework for On-the-Fly Anatomical Location Matching between Longitudinal Studies}%
\titlerunning{Point Matching}
\author{Halid Ziya Yerebakan \and Yoshihisa Shinagawa \and Mahesh Ranganath \and Simon Allen-Raffl \and Gerardo Hermosillo Valadez }
%
\authorrunning{Yerebakan et. al.}

\institute{Siemens Medical Solutions, Malvern, USA \email{halid.yerebakan@siemens-healthineers.com}}

\maketitle

\begin{abstract}
    We propose a method to match anatomical locations between pairs of medical images in longitudinal comparisons. The matching is made possible by computing a descriptor of the query point in a source image based on a hierarchical sparse sampling of image intensities that encode the location information. Then, a hierarchical search operation finds the corresponding point with the most similar descriptor in the target image. This simple yet powerful strategy reduces the computational time of mapping points to a millisecond scale on a single CPU. Thus, radiologists can compare similar anatomical locations in near real-time without requiring extra architectural costs for precomputing or storing deformation fields from registrations. Our algorithm does not require prior training, resampling, segmentation, or affine transformation steps. We have tested our algorithm on the recently published Deep Lesion Tracking dataset annotations. We observed more accurate matching compared to Deep Lesion Tracker while being 24 times faster than the most precise algorithm reported therein. We also investigated the matching accuracy on CT and MR modalities and compared the proposed algorithm's accuracy against ground truth consolidated from multiple radiologists. 
    
    
\end{abstract}


\section{Introduction}
\label{sec:introduction}
	For many medical image findings, such as lesions, it is essential to assess the progress over time. Thus, comparison between different studies on the medical history of patients is often an indispensable step. However, there is no direct correspondence between the coordinate systems across 3D volumes in different time points due to various scanning conditions. Thus, radiologists must manually navigate to the same anatomical locations in each comparison study. 




%

    The mainstream approach for aligning coordinate systems across images is registration. Traditional registration algorithms suffer from computational time since every image pair needs to optimize the deformation field on runtime. Deep learning approaches have improved the computational time of iterative deformable registration methods by learning deformation estimation functions from given pair of images \cite{balakrishnan2019voxelmorph,huang2021coarse,guo2019multi,hasenstab2021cnn,mok2020large}. However, the learned network would be optimal only for the dataset types that are present in training. Thus, domain shift adaptation and transfer learning approaches are active research areas in deep learning based medical image registration  \cite{varoquaux2022machine,tang2022self}. 
    
    Full registration is unnecessary for many navigational point matching tasks in routine reading \cite{cai2021deep,samtmi}. Only corresponding points can be found to reduce computational time. Yan et al. and Cai et al. formulated the problem as a tracking problem \cite{samtmi,cai2021deep}. They developed an unsupervised learning algorithm for embeddings of anatomical locations in medical images. They used coarse-to-fine embedding levels to combine local and global information. Thanks to these embedding descriptors, finding the corresponding location is reduced to a single convolution operation, which could be computed in 0.23s using GPU. Liu et al. have extended these descriptors to full registration later\cite{liu2021same}. The limited availability of GPUs on runtime and limited generalization of deep models beyond trained body regions or modalities remains a challenge for these models. In the literature, there are other unsupervised/self-supervised/handcrafted descriptor approaches that could be used within similar search framework \cite{blendowski2019learn,heinrich2012mind,chaitanya2020contrastive}. 
	
	Our point matching algorithm is a descriptor search framework similar to this line of work \cite{samtmi,cai2021deep}. However, unlike deep learning models, our descriptor is sampling based descriptor, which does not need optimization for training and GPU hardware acceleration. Also, unlike the traditional image matching and tracking methods, our feature representation is more efficient to create, it has a large global receptive field, and it is easy to scale for hierarchical search. \cite{grauman2005pyramid,bay2008speeded,karami2017image}. Our method is independent of voxel scaling and modality thanks to world coordinates in medical images. Thus, it has a negligible memory footprint in runtime without the need for preprocessing like affine transformation or resampling.

 	Our algorithm finds 83.6\%  of points within 10mm (or radius for small findings)  on the public Deep Lesion Tracking dataset \footnote{https://github.com/JimmyCai91/DLT} while having runtimes below 200ms per pair without additional hardware. Also, we have a 1.685 mm median distance error on the mixed modality (CT, MR) dataset, which indicates the general applicability of our method. Our algorithm exhibits higher accuracy as compared to the state-of-the-art machine learning methods while being an order of magnitude faster on runtime.
	
\section{Method} 

    To compare specific locations in follow-up studies, we can set up a function that establishes correspondence in pairs of images. This method is more efficient than fully registering images when only a few points are queried. For example, we can compare a given descriptor in the source image location against every voxel descriptor in the target image. However, this process can still be computationally expensive, especially within 3D volumes. To address this, we've implemented a hierarchical approach in the descriptor formulation and the search processes. 
    
	
\subsection{Descriptor Computation}

\begin{figure}
	\centering
	\includegraphics[width=0.4\linewidth]{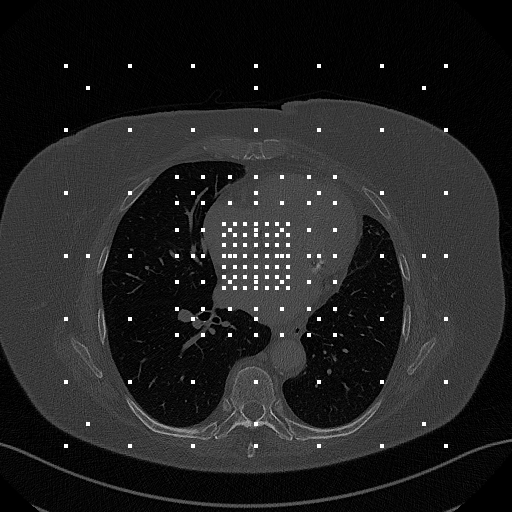} \includegraphics[width=0.4\linewidth]{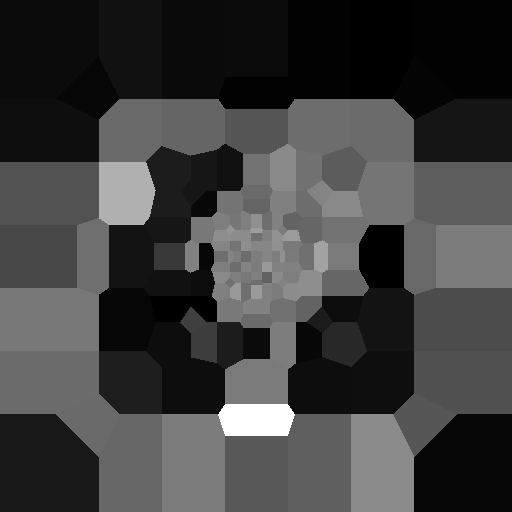}
	\caption{Descriptors are computed based on fixed sampling offsets for a query location (left). Decoding of the descriptor with nearest neighbor intensities  (right).}
	\label{fig:sampling}
\end{figure}


     The central component of any descriptor search approach is designing a good descriptor that produces similar vectors in the same anatomical regions and distinct vectors for different anatomical locations. Distances to neighbor organs are as important as the intensity values at the center to describe the location of the query. Unlike natural images, medical images have accurate spatial dimensions thanks to the world reference frame information included in the image header meta-data, which is the base for medical abnormality assessments. It is especially more stable in longitudinal studies where images are from the same patient. Thus, sampling intensity values with the same spatial offsets have a strong correlation in similar anatomical locations within comparison images.


	 
	
    We first create a sampling model where we define the location offsets in actual mm distances. We prefer more samples in the center and fewer samples in the peripheral regions to reduce descriptor size. This would create a balance between the precision of the location estimation and robustness thanks to the large global field of view. This type of pyramid hierarchy is one of the key ideas in computer vision, validated in many forms for various tasks. In our experiments, we have used 8mm, 20mm, 48mm, and  128mm distance grids with a grid size of 7x7x7 in each resolution. As a result, this model creates 1372-dimensional vector descriptors. The maximum value of offsets is 384mm in this setting, which is sufficient to cover most of the body sizes in medical images. A 2D projection visualization of sampling points is shown in Figure \ref{fig:sampling}. This sampling model is scaled down by a factor of 1/2 on each subsequent level of search. In the latest 5th level, these sizes become 0.5mm, 1.25mm, 3mm, and 8mm in terms of distances which are multiples of typical slice thickness values in CT and MR images. These hyperparameters could be changed according to desired imaging modality. However, we have used the same hyperparameters in all experiments.  

    After defining the offset model in millimeters, we compute voxel offset according to voxel spacing in the volume. This step will eliminate the requirement of resampling images into the canonical voxel spacings. Instead of adapting the images, we have adapted the offsets of the sampling model accordingly once per image and search level. Thus, the descriptor generation becomes a memory lookup operation of pre-computed offset locations for any given point in the image with almost zero computation and negligible storage, which is much more efficient than any other traditional or deep learning feature extraction.  If the offset locations are outside the image volume, they are given the value of 0 for the corresponding dimension. Thus, all descriptors are in the same vector space. 
  
	 

    The descriptors are encoded versions of the image since the sampler uses the intensity values directly. Thus, it is possible to reconstruct the image back from the descriptor by using the sampled nearest neighbor intensity values of each voxel location. We demonstrated the center slice of an exemplary reconstructed image from the center point descriptor, as shown in Figure \ref{fig:sampling}. Our method is sensitive to location since the sampling points behave like distance sensors. Thus, the variance in the visualization is higher than in basic resampling methods. But, this is desirable since translation invariance is harmful for encoding location.

\subsection{Similarity}

Various similarity measurements are applicable to the proposed approach. However, CT and MR modalities differ from each other in terms of similarity measures. In CT images, intensities are often close to each other due to Hounsfield unit standardization. Thus, Euclidean and cosine similarities are valid choices. We used full vector dimension into cosine similarity which includes the edge information. Intensities and contrasts are variable in the MR modality according to acquisition parameters. In this case, more intensity invariant metric, such as mutual information, is necessary. In our experiments, we obtained better results by combining cosine and mutual information similarities for both modalities.


Mutual information is calculated by taking the joint histogram probability of descriptors $p(x,y)$ of the bin of x and y. The ranges between maximum and minimum intensity values are divided into 16 histogram bins, and the whole histogram is normalized into joint probabilities $p(x,y)$. Then, marginal $p(x)$ and $p(y)$ probabilities could be obtained by adding $p(x,y)$ among rows and columns. Also, sampling offsets outside of the imaging region (which was set to 0) are excluded from mutual information calculation since otherwise, it would distort histogram bins. Finally, the mutual information between two descriptors could be computed as given in the equation \eqref{eq} where $K$ is the number of bins. 

\begin{equation}\sum_{x=1}^{K}\sum_{y=1}^{K} p(x,y)log(p(x,y)/(p(x)p(y))) \label{eq}\end{equation}

\subsection{Hierarchical Search}

\begin{figure}
  \begin{subfigure}{0.5\linewidth}
    \centering
    \includegraphics[trim={1cm 1cm 1cm 1cm},clip,width=\linewidth]{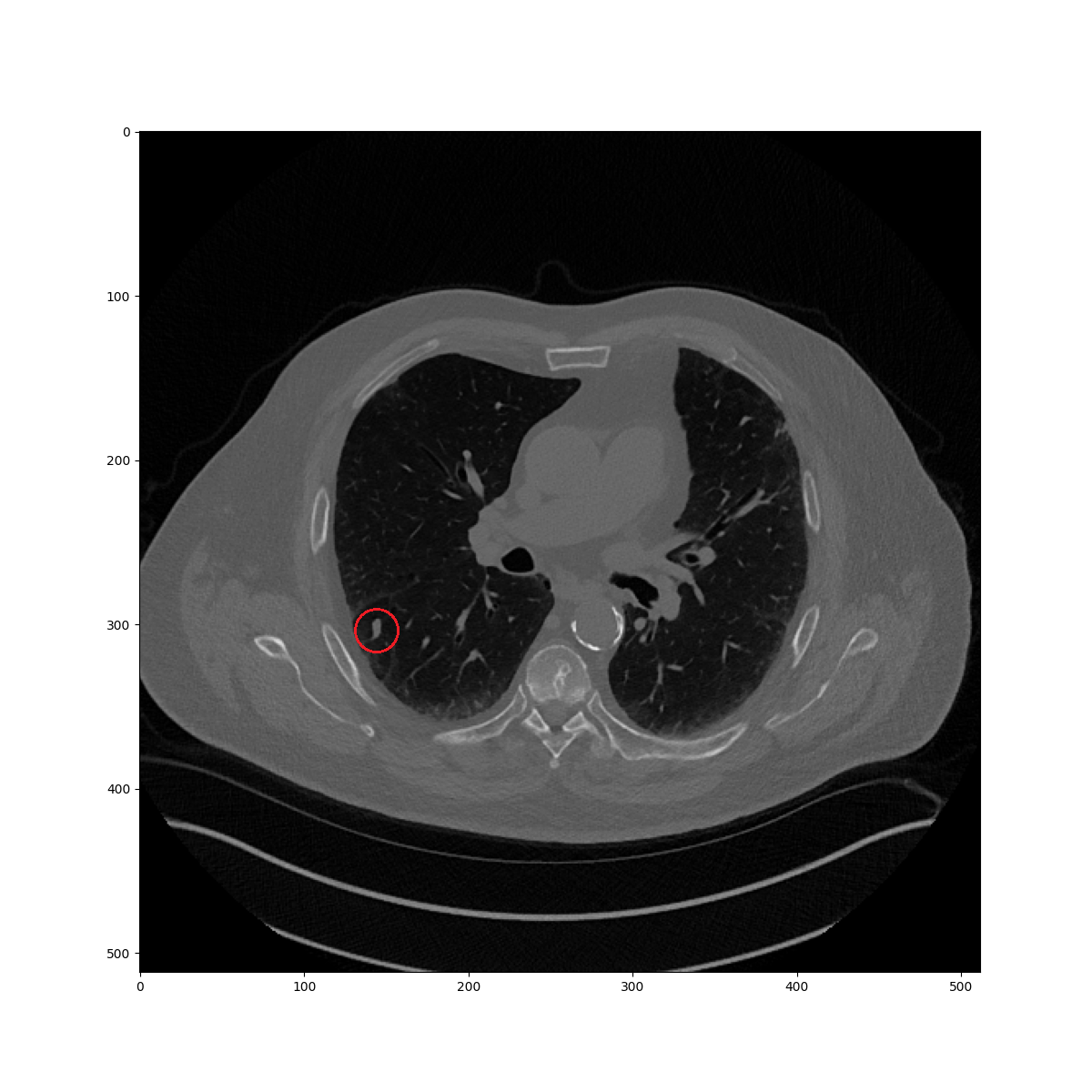}
    
    \caption{Source image query location}
    \label{fig:current}

  \end{subfigure}%
  \begin{subfigure}{0.5\linewidth}
    \centering
    \includegraphics[trim={1cm 1cm 1cm 1cm},clip,width=\linewidth]{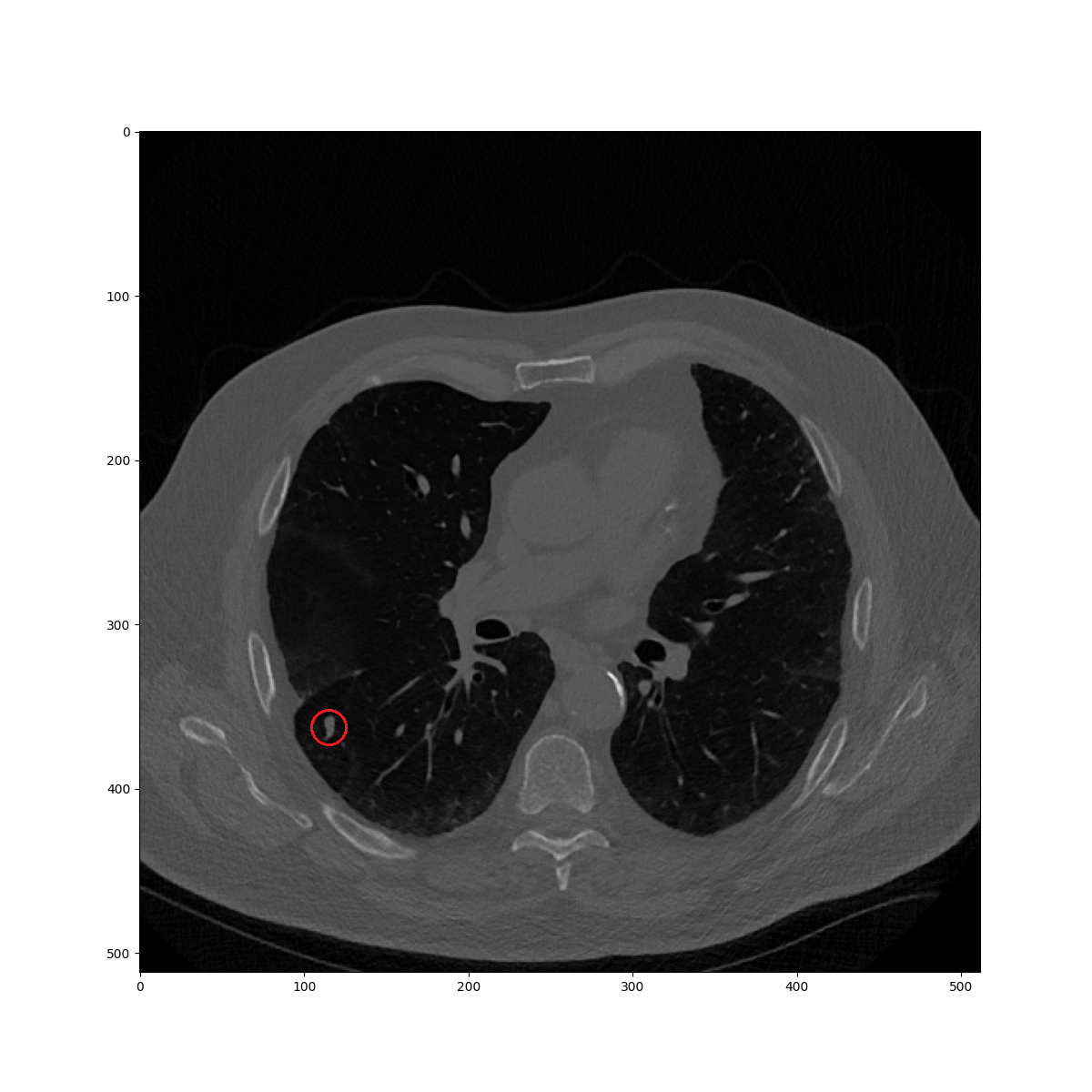}

\caption{Target image ground truth location}
 \label{fig:prior}
  \end{subfigure}
  
  \begin{subfigure}{0.5\linewidth}
    \centering
    \includegraphics[trim={1cm 1cm 1cm 1cm},clip,width=\linewidth]{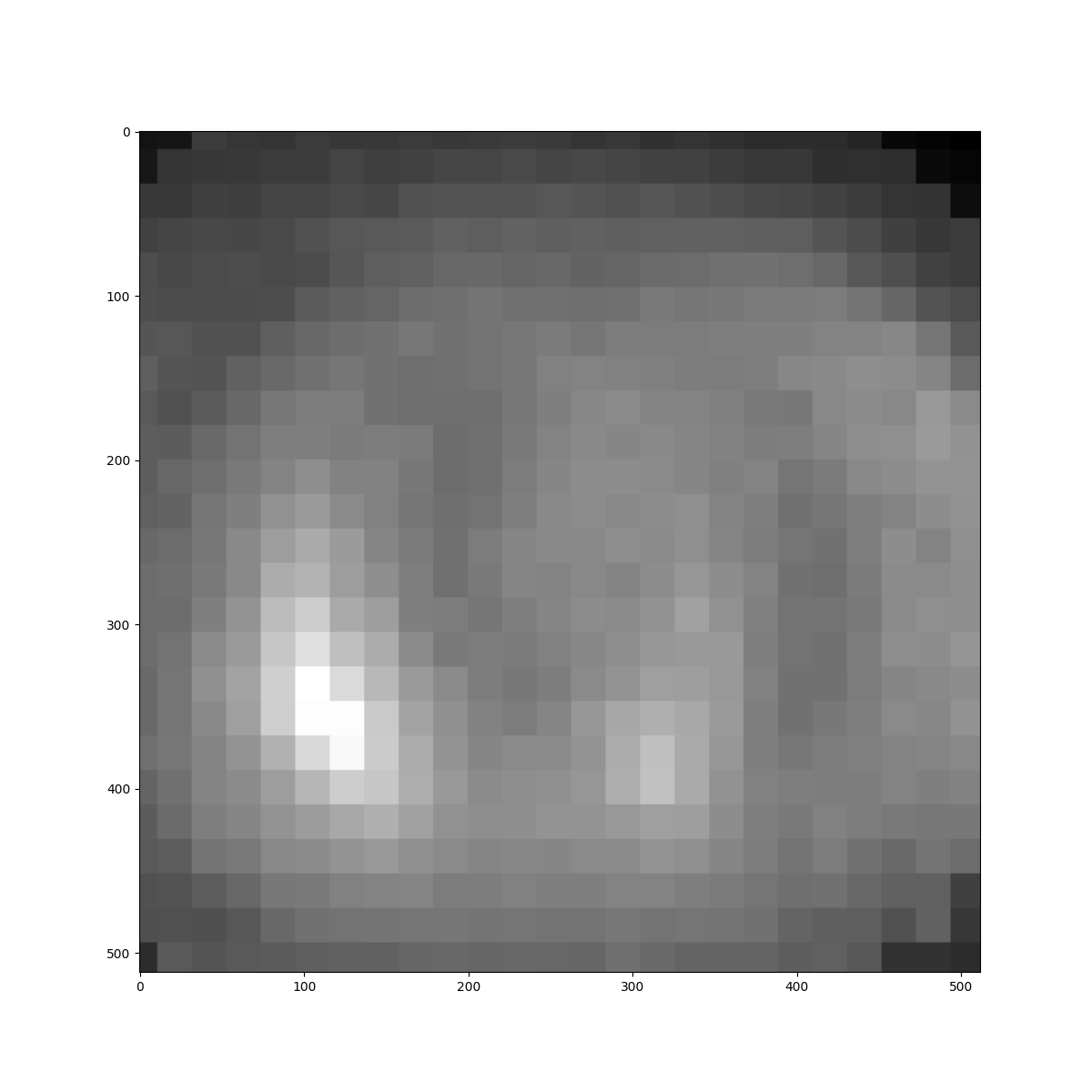}
\caption{Coarse similarity map on target image}
\label{fig:coarse}
  \end{subfigure}%
  \begin{subfigure}{0.5\linewidth}
    \centering
    \includegraphics[trim={1cm 1cm 1cm 1cm},clip,width=\linewidth]{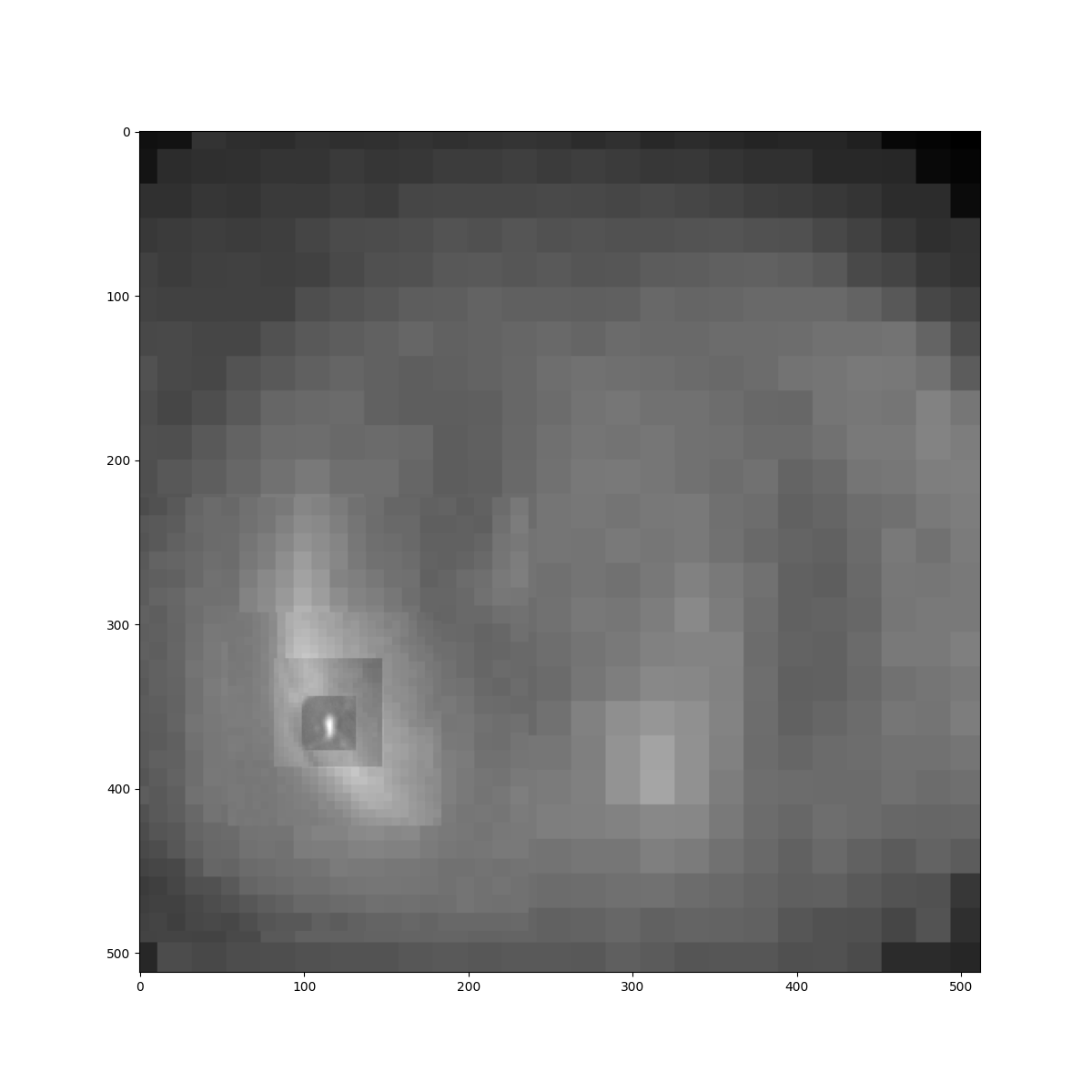}

\caption{Hierarchical search on target image}
\label{fig:fine}
  \end{subfigure}
    \begin{subfigure}{0.5\linewidth}
    \centering
    \includegraphics[width=\linewidth,trim={20pt 20pt 20pt 20pt},clip]{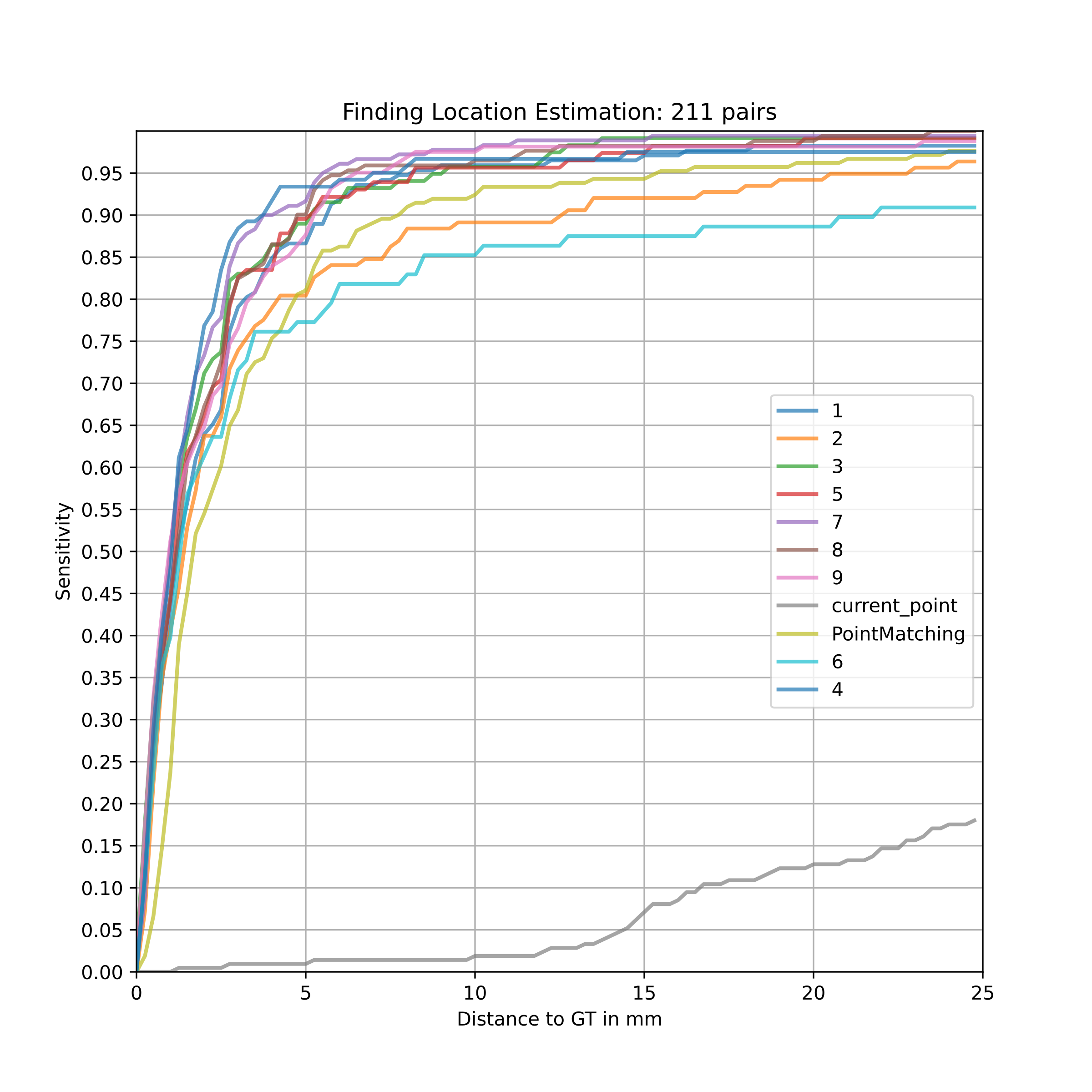}
	\caption{FROC on CT, MR dataset}
	\label{fig:humanannotators}
  \end{subfigure}%
  \begin{subfigure}{0.5\linewidth}
    \centering
    \includegraphics[width=\linewidth,trim={20pt 20pt 20pt 20pt},clip]{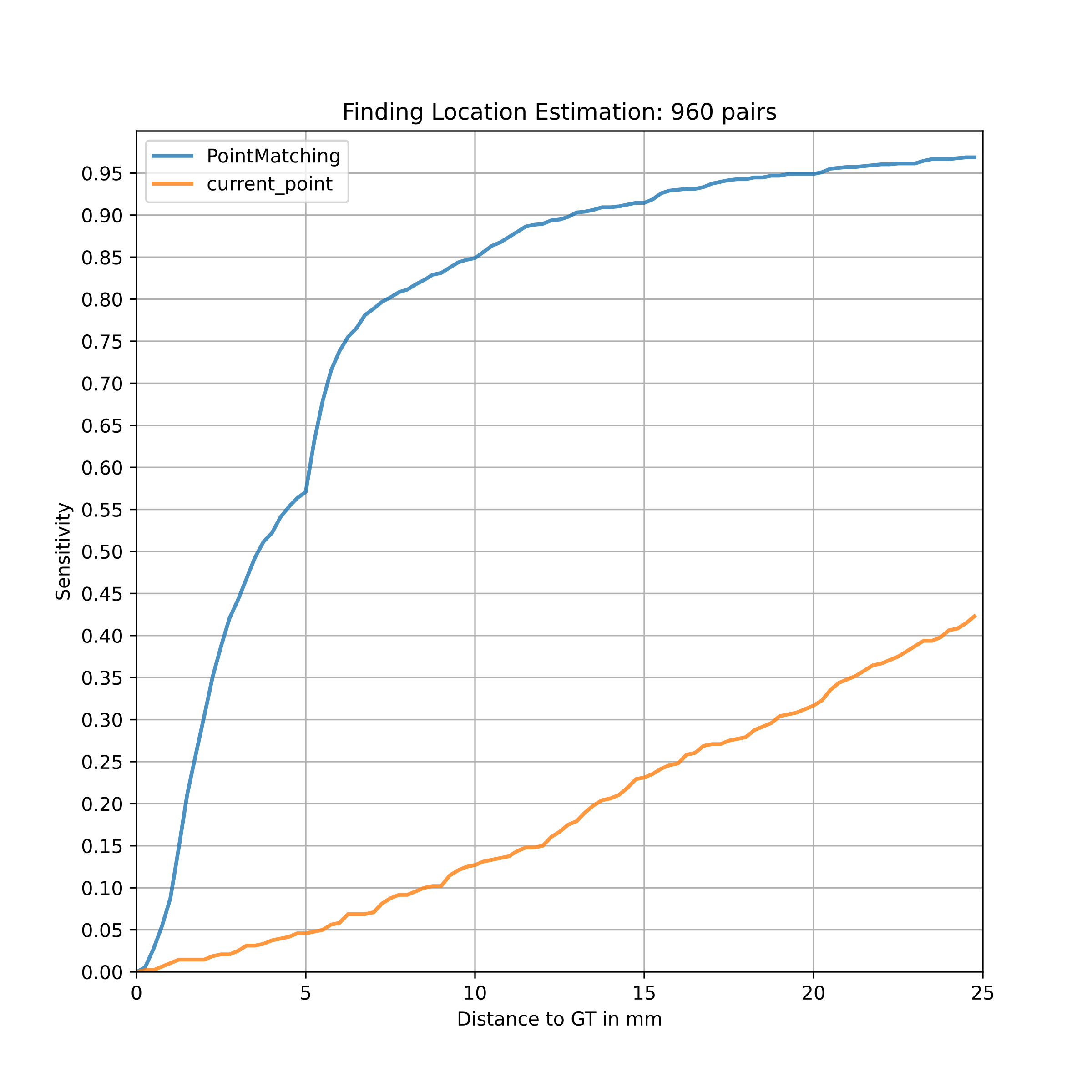}
    \caption{FROC on Deep Lesion Tracking dataset}
    \label{fig:deeplesion}
  \end{subfigure}
  
  \caption{Hierarchical Search and Results on Multiple Time Points}
\end{figure}

	Computing the similarity for every voxel in the target image is computationally expensive. Also, the base sampling model in the original scale is already coarse for descriptor computation which would create similar results on nearby voxels at the top level. Thus, some voxels could be skipped during the search. Based on these facts, we have utilized a coarse-to-fine search strategy to find the corresponding location in the target image. 
	
	In the first level, the method starts with a 16 mm equidistant grid. The selected similarity metric is computed for every location in that grid. Then the maximum similarity location is selected to advance the granularity of the search around that location. In the next level, the original sampling model and grid size are scaled down by 1/2. We have applied five scaling levels that result in the finest level of search as the 1mm grid. Additionally, search space is reduced to avoid mapping to distant regions from the previous level estimate and to improve computational efficiency. The search box size is set to 96mm after doing a whole volume search, and it is reduced to 12mm in the final level. 
	
    In Figure \ref{fig:current}, there is a lung nodule at the point of interest of the current CT volume as an example of query point. The task is finding the corresponding location in the prior CT volume as shown in \ref{fig:prior}. In the first level, similarities are computed in every location of the 16mm grid which could be seen as a low-resolution heatmap as in \ref{fig:coarse}. Then, until the fifth level, the maximum similarity location is used to continue the search in the next level while reducing the size of the search region. The resulting heatmap is given in Figure \ref{fig:fine}. The similarity in the finest level precisely describes the nodule location. There are two different characteristics as compared to the deep learning-based descriptor search approach \cite{samtmi}. First, the finest similarity is not computed on the whole image but on a specific location conditionally on the previous maximum. Thus, our algorithm is computationally faster, even on the CPU. Also, the descriptor is based on a predefined sampling model, which does not require optimization in training time or any additional computation for feature extraction in the runtime.

    
	
	
	
    Searching in each level is straightforward to parallelize since each similarity calculation is independent with a read-only lookup operation to get candidate descriptors in the target image. Thus, we can compute the different candidate point locations in separate threads. Then, we can select the maximum similarity location after completing all grid points. This approach has massive parallelization potential. However, in practice, CPU hardware has a limited number of threads for parallelization. Thus, we have applied parallelization on the slice level to reduce the overhead of thread creation. 

\section{Experiments}


\subsection{Comparison Study}

In our first experiment, we compared the proposed method with respect to recently published state-of-the-art results. We utilized Deep Lesion Tracking dataset with testing annotations published by \cite{cai2021deep}. In our case, however, we did not utilize the training set since training was not necessary. Also, it is worth to note that deep lesion dataset has limited number of slices along z axis which makes it more challenging as compared to regular CT scans. 

We have used similar slice sorting steps for loading the image as given in the auxiliary script. Additionally, we have added value 1024 to images to represent them in unsigned integer values due to our imaging environment. The testing set contains 480x2 lesion pairs, including both directions from pairs of time points. All images are in the CT modality for this dataset. 



\begin{table*}
\centering
\caption{Deep Lesion Tracking Dataset Results}
\setlength{\tabcolsep}{3pt}
\begin{tabular}{|p{70pt}|p{70pt}|p{70pt}|p{70pt}|}
\hline
Method &  CPM@10mm &  Average (mm) &  Speed(s) \\
\hline
No Registration & 12 &  28.71 & - \\
Affine &  48.33 &  11.2 &  1.82 \\
VoxelMorph & 49.90  &  10.9 &  0.46 \\
DEEDS &  71.88 &  7.4 &  15.3 \\
DLT &  78.85 & 7.0 & 3.58 \\
PointMatching &  {\bf 83.6 } & {\bf 5.94    }& {\bf 0.149} \\
\hline
\end{tabular}
\label{table:comparison}
\end{table*}

We have used the same metrics in our evaluation: Euclidean distance and sensitivity at adaptive distance threshold min(r, 10mm), where r represents the radius of the finding in comparison as in \cite{cai2021deep}. Our results are shown in Table \ref{table:comparison}. 


The proposed Point Matching finds closer estimate points with a mean of 5.94mm in target ground truth compared to other registration or descriptor search methods. We have illustrated the sensitivity at different distance thresholds in an FROC as in Figure \ref{fig:deeplesion}, which represents the cumulative distribution of distances. There is a discontinuity of slope of graph  at 5mm due to the dominant slice thickness of 5mm. So, if the estimation is one slice off in those cases, the distance would be greater than 5mm.


The related work reported speed results based on GPU hardware. Our point matching algorithm is faster (149 ms) than all GPU accelerated registration algorithms, even using a single i9-7900x CPU. We have used C++ and OpenMPI for the implementation of the algorithm with 12 threads for parallelization. 




\subsection{Mixed Modality Dataset}

We have evaluated our method additionally on an in-house study dataset containing multi-time point CT and MR modalities. Our dataset contains aortic aneurysms, intracranial aneurysms (ICA), enlarged lymph nodes, kidney lesions, meningioma, and pulmonary nodule pathologies. In this study, the annotations come from multiple annotators. Radiologists were presented with pairs of images along with a description of a predefined finding in the current studies. They were asked to find the corresponding locations in the prior studies. 

The measurements are consolidated as ground truth by up to 9 different radiologists by taking the median of their annotations. We have used a series with more than 3 annotators to compare different time points. Overall, 211 pairs of series were selected with this criteria.

We have estimated the corresponding prior locations of the findings with our point matching algorithm compared with expert annotations. We have illustrated the change in sensitivity with different distance thresholds in Figure \ref{fig:humanannotators}. Each annotator is labeled with a number in this figure. Annotators have better localization below the 5mm. However, due to some annotation disagreement in a few cases, there are two annotators below the automated algorithm in a larger distance range. Notably, our algorithm is very close to an average radiologist annotation robustness at 25mm. 


\begin{table}
\centering
\caption{Mixed modality (CT, MR) dataset clinical findings' median errors (mm)}
\setlength{\tabcolsep}{3pt}
\begin{tabular}{|p{70pt}|p{50pt}|p{50pt}|p{50pt}|p{50pt}|p{50pt}|}
\hline
 Name  & Patients & Pairs &  Type & Ours & Expert  \\
\hline
Aneurysm & 15 & 44 & CT & 4.24 & 2.86 \\
Lung Nodules & 15 & 37 & CT & 1.14 & 0.45 \\
Kidney Lesions & 15 & 42 & CT & 3.08 & 1.75 \\
ICA & 15 & 33 & MR & 0.92 & 0.59 \\
Lymph Nodes & 10 & 27 & CT & 3.29 & 2.67 \\
Meningioma & 15 & 28 & MR & 1.62 & 1.3 \\
All & 85 & 211 & * & 1.685 & 1.11 \\
\hline
\end{tabular}
\label{table:cases}
\end{table}

We have also investigated medians of location estimations of individual pathologies and modalities. The median values of estimations and expert annotations are given in millimeters in Table \ref{table:cases}. Lymph nodes and Kidney lesions are more difficult due to less contrast and more variation in abdominal regions. Also, more variation in the center location of findings is present in aortic aneurysms due to larger measurement sizes. Lung nodules are very precise thanks to the reliable chest wall contrast. Similarly, brain MRI studies present good matching performance with the proposed algorithm. Overall algorithm performance is close to the expert annotators with a difference of only 0.6 mm in the medians.

\section{Conclusion}

	We present a simple yet effective method of mapping corresponding locations in a pair of volumetric medical images. Unlike landmarking methods, our method allows arbitrary locations in the source image to be mapped to the target image. Unlike the registration algorithms, it does not map all the points apriori without losing its non-rigid mapping ability. Unlike the deep learning approaches, there is no computation needed to calculate descriptors in the runtime as well as no optimization in training time. Instead, it populates an efficient descriptor with memory lookups in the source image location and finds the corresponding location with the maximum similarity match in the target image with a hierarchical search. Thus, no resampling, initialization,  or affine transformation is necessary for the preprocessing. It runs favorably on CPUs while better location estimates than state-of-the-art according to annotations in the public Deep Lesion Benchmark dataset. Additionally, our method works on both CT and MR modalities on different body parts with location estimates comparable to expert annotators. 

\nocite{weikert2023reduction}
\bibliographystyle{splncs04}
\bibliography{pointmatching}

\begin{thebibliography}{10}
\providecommand{\url}[1]{\texttt{#1}}
\providecommand{\urlprefix}{URL }
\providecommand{\doi}[1]{https://doi.org/#1}

\bibitem{balakrishnan2019voxelmorph}
Balakrishnan, G., Zhao, A., Sabuncu, M.R., Guttag, J., Dalca, A.V.: Voxelmorph:
  a learning framework for deformable medical image registration. IEEE
  transactions on medical imaging  \textbf{38}(8),  1788--1800 (2019)

\bibitem{bay2008speeded}
Bay, H., Ess, A., Tuytelaars, T., Van~Gool, L.: Speeded-up robust features
  (surf). Computer vision and image understanding  \textbf{110}(3),  346--359
  (2008)

\bibitem{blendowski2019learn}
Blendowski, M., Nickisch, H., Heinrich, M.P.: How to learn from unlabeled
  volume data: Self-supervised 3d context feature learning. In: International
  Conference on Medical Image Computing and Computer-Assisted Intervention. pp.
  649--657. Springer (2019)

\bibitem{cai2021deep}
Cai, J., Tang, Y., Yan, K., Harrison, A.P., Xiao, J., Lin, G., Lu, L.: Deep
  lesion tracker: Monitoring lesions in 4d longitudinal imaging studies. In:
  Proceedings of the IEEE/CVF Conference on Computer Vision and Pattern
  Recognition. pp. 15159--15169 (2021)

\bibitem{chaitanya2020contrastive}
{Chaitanya}, K., {Erdil}, E., {Karani}, N., {Konukoglu}, E.: Contrastive
  learning of global and local features for medical image segmentation with
  limited annotations. In: Advances in Neural Information Processing Systems 33
  pre-proceedings (NeurIPS 2020). vol.~33, pp. 12546--12558 (2020)

\bibitem{grauman2005pyramid}
Grauman, K., Darrell, T.: The pyramid match kernel: Discriminative
  classification with sets of image features. In: Tenth IEEE International
  Conference on Computer Vision (ICCV'05) Volume 1. vol.~2, pp. 1458--1465.
  IEEE (2005)

\bibitem{guo2019multi}
Guo, C.K.: Multi-modal image registration with unsupervised deep learning.
  Ph.D. thesis, Massachusetts Institute of Technology (2019)

\bibitem{hasenstab2021cnn}
Hasenstab, K.A., Tabalon, J., Yuan, N., Retson, T., Hsiao, A.: Cnn-based
  deformable registration facilitates fast and accurate air trapping
  measurements on inspiratory-expiratory ct. Radiology: Artificial Intelligence
  p. e210211 (2021)

\bibitem{heinrich2012mind}
Heinrich, M.P., Jenkinson, M., Bhushan, M., Matin, T., Gleeson, F.V., Brady,
  M., Schnabel, J.A.: Mind: Modality independent neighbourhood descriptor for
  multi-modal deformable registration. Medical image analysis  \textbf{16}(7),
  1423--1435 (2012)

\bibitem{huang2021coarse}
Huang, W., Yang, H., Liu, X., Li, C., Zhang, I., Wang, R., Zheng, H., Wang, S.:
  A coarse-to-fine deformable transformation framework for unsupervised
  multi-contrast mr image registration with dual consistency constraint. IEEE
  Transactions on Medical Imaging  (2021)

\bibitem{karami2017image}
Karami, E., Prasad, S., Shehata, M.: Image matching using sift, surf, brief and
  orb: performance comparison for distorted images. arXiv preprint
  arXiv:1710.02726  (2017)

\bibitem{liu2021same}
Liu, F., Yan, K., Harrison, A.P., Guo, D., Lu, L., Yuille, A.L., Huang, L.,
  Xie, G., Xiao, J., Ye, X., et~al.: Same: Deformable image registration based
  on self-supervised anatomical embeddings. In: International Conference on
  Medical Image Computing and Computer-Assisted Intervention. pp. 87--97.
  Springer (2021)

\bibitem{mok2020large}
Mok, T.C., Chung, A.C.: Large deformation diffeomorphic image registration with
  laplacian pyramid networks. In: International Conference on Medical Image
  Computing and Computer-Assisted Intervention. pp. 211--221. Springer (2020)

\bibitem{tang2022self}
Tang, Y., Yang, D., Li, W., Roth, H.R., Landman, B., Xu, D., Nath, V.,
  Hatamizadeh, A.: Self-supervised pre-training of swin transformers for 3d
  medical image analysis. In: Proceedings of the IEEE/CVF Conference on
  Computer Vision and Pattern Recognition. pp. 20730--20740 (2022)

\bibitem{varoquaux2022machine}
Varoquaux, G., Cheplygina, V.: Machine learning for medical imaging:
  methodological failures and recommendations for the future. npj Digital
  Medicine  \textbf{5}(1), ~1--8 (2022)

\bibitem{weikert2023reduction}
Weikert, T., Litt, H.I., Moore, W.H., Abed, M., Azour, L., Noor, A.M., Friebe,
  L., Linna, N., Yerebakan, H.Z., Shinagawa, Y., et~al.: Reduction in
  radiologist interpretation time of serial ct and mr imaging findings with
  deep learning identification of relevant priors, series and finding
  locations. Academic Radiology  (2023)

\bibitem{samtmi}
Yan, K., Cai, J., Jin, D., Miao, S., Guo, D., Harrison, A.P., Tang, Y., Xiao,
  J., Lu, J., Lu, L.: Sam: Self-supervised learning of pixel-wise anatomical
  embeddings in radiological images. IEEE Transactions on Medical Imaging
  pp.~1--1 (2022). \doi{10.1109/TMI.2022.3169003}

\end{thebibliography}

\end{document}